\documentclass{sf2a-conf}
\usepackage{graphicx}

\begin{document}
\TitreGlobal{SF2A 2008}
%%-----------------------------
%%      the top matter
%%-----------------------------
\title{Neutrino detection of transient sources with optical follow-up observations}
\author{Dornic D.} \address{CPPM, Faculte des Sciences de Luminy, 163 avenue de Luminy, 13288
Marseille Cedex09, France}
\author{Basa S.} \address{LAM, BP8, Traverse du Siphon, 133376 Marseille Cedex 12, France}
\author{Brunner J.$^1$} 
\author{Busto J.$^1$}
\author{Boer M.} \address{OHP, 04870 Saint Michel L'Observatoire, France}
\author{Klotz A.$^{3,}$} \address{CESR, Observatoire Midi-Pyr\'en\'ees, CNRS, Universit\'e de Toulouse, BP 4346, 31028 Toulouse Cedex 04, France }
\author{Escoffier S.$^1$}
\author{Gendre B.$^2$} 
\author{Le Van Suu A.$^3$} 
\author{Mazure A.$^2$}
\author{Atteia J.L.} \address{LATT, Observatoire Midi-Pyr\'en\'ees, CNRS, Universit\'e de Toulouse, 14 avenue E. Berlin, 31400 Toulouse, France}
 \author{Vallage B.} \address{CEA-IRFU, centre de Saclay, 91191 Gif-Sur-Yvette, France}
\runningtitle{}
\setcounter{page}{237} % Keep this line, even if the page will be settled afterwards..

%\author{Dornic D.} 
%\author{Brunner J.} 
%\author{Busto J.}
%\author{Escoffier S.}
%\author{Vallage B.} 
%\author{Basa S.} 
%\author{Gendre B.} 
%\author{Mazure A.} 
%\author{Atteia JL.} 
%\author{Klotz A.} 
%\author{Boer M.} 
%\author{Le Van Suu.} 
% Repeat the authors here, this will help to make the final index

\maketitle
\begin{abstract}
The ANTARES telescope has the opportunity to detect transient neutrino sources, such
as gamma-ray bursts (GRBs), core-collapse supernovae (SNe), flares of active galactic nuclei
(AGNs)... To enhance the sensitivity to these sources, we are developing a new detection
method based on the observation of neutrino bursts followed by an optical detection. The ANTARES
Collaboration is implementing a fast on-line event reconstruction with a good angular resolution.
These characteristics allow to trigger an optical telescope network in order to identify the
nature of the neutrinos (and high energy cosmic-rays) sources. This follow-up can be done
with a network of small automatic telescopes and required a small observation time. An
optical follow-up of special events, such as neutrino doublets in coincidence in time and space
or single neutrino having a very high energy, would not only give access to the nature of the
sources but also improve the sensitivity for neutrino detection from SNe or GRBs.
\end{abstract}
%
%%-----------------------------
%%      your text
%%-----------------------------
\section{Introduction}
The ANTARES neutrino telescope [{\cite{BAntares}}] is located 40 km off shore Toulon, in the South French coast, at about 
2500\,m below sea level. The complete detector is composed of 12 lines, each including 75 photomultipliers 
on 25 storeys, which are the sensitive elements. Data taking started in 2006 with the operation of the 
first line of the detector. The construction of the 12 line detector was completed in May 2008. The main goal 
of the experiment is to detect high energy muon induced by neutrino interaction in the vicinity of the 
detector. 

Among all the possible astrophysical sources, transients offer one of the most promising perspectives 
for the detection of cosmic neutrinos thank to the almost background free search. The detection of these 
neutrinos would be the only direct probe of hadronic accelerations and so, the discovery of the ultra 
high energy cosmic ray sources without ambiguity.

In this paper, we discuss the different strategies implemented in ANTARES for the transient sources 
detection.

%*************************************************************************************************************************
%*************************************************************************************************************************
%*************************************************************************************************************************

\section{Transient sources detection strategies}
\label{section1}
To detect transient sources, two different methods can be used [{\cite{BBasa}}]. The first one is based on the search for 
neutrino candidates in conjunction with an accurate timing and positional information provided by an external 
source: the triggered search. The second one is based on the search for "special neutrino events" coming from 
the same position within a given time window: the rolling search.

\subsection{The triggered search}
\label{subsection1}
Classically, GRBs or flare of AGNs are detected by gamma-ray satellites which deliver an 
alert to the Gamma-ray bursts Coordinates Network (GCN [{\cite{BGCN}}]). The characteristics of this alert are then 
distributed to the other observatories. The small difference in arrival time and position expected between 
photons and neutrinos allows a very efficient detection even with a low energy threshold due to the very low expected 
background. This method has been implemented in ANTARES mainly for the GRBs detection since the 
end of 2006. Today, the alerts are primarily provided by the Swift [{\cite{BSwift}}] and the Integral 
[{\cite{BIntegral}}] satellites. Data triggered by more than 250 alerts have been stored up to now.

Due to the very low background rate, even the detection of a small number of neutrinos correlated with GRBs 
could set a discovery. But, due to the relatively small field of view of the gamma-ray satellites (for example, 
Swift has a 1.4\,sr field of view), only a small fraction of the existing bursts are triggered. Moreover, the choked 
GRBs without photons counterpart can not be detected by this method.

\subsection{The rolling search}
\label{subsection2}
This second method, originally proposed by Kowalski and Mohr [{\cite{BKowalski}}], consists on the detection of a burst of neutrinos 
in temporal and directional coincidence. Applied to ANTARES, the detection of a doublet of neutrinos is almost 
statistically significant. Indeed, the number of doublet due to atmospheric neutrinos is of the order of 0.05 per year 
when a temporal window of 900\,s and a directional one of 3\,$^\circ$ x 3\,$^\circ$ are defined.

It is also possible to search for single cosmic neutrino events by requiring that the reconstructed muon energy is 
higher than a given energy threshold (typically above a few tens of TeV if a Waxman$-$Bahcall flux is considered). 
This high threshold reduces significantly the atmospheric neutrino background [{\cite{BDornic}}].

In contrary to the current gamma-ray observatories, a neutrino telescope covers at least a half hemisphere if 
only up-going events are analyzed and even $4\pi$\,sr if down-going events are considered. When the neutrino 
telescope is running, this method is around 100\% efficient. Moreover, this method requires no hypothesis on the 
period during which the neutrinos are emitted with respect to the gamma flash, a parameter not really 
constrained by the different models. More importantly no assumption is made on the nature of the source and the 
mechanisms occurring inside.

The main drawback of the rolling search is that a detection is not automatically associated to an astronomical 
source. To overcome this problem, it is fundamental to organize a complementary follow-up program. This program 
can be done with a small optical telescope network. The observation of any transient sources will require a 
quasi real-time analysis and a very good angular precision. It will be described in detail in section 4.

%*************************************************************************************************************************
%*************************************************************************************************************************
%*************************************************************************************************************************

\section{The ANTARES neutrino triggers}
\label{section2}
ANTARES is implementing a new on-line event reconstruction (named BBfit). This analysis strategy contains a very 
efficient trigger based on local clusters of photomultiplier hits and a simple event reconstruction. The two main 
advantages are a very fast analysis (between 5 and 10\,ms per event) and a good angular resolution. The minimal 
condition for an event to be reconstructed is to contain a minimum of six storeys triggered on at least two lines. 
To select a high purity sample of up-going neutrino candidates, one quality cut is applied to the result of the $\chi^2$ 
minimisation of the muon track reconstruction based on the measured time and amplitude of the hits. In order to obtain a 
fast answer, the on-line reconstruction does not use the dynamic reconstructed geometry of the detector lines. This has the 
consequence that the angular resolution is degraded with respect to the one obtained with the standard off-line ANTARES 
reconstruction (of about 0.2 - 0.3\,$^\circ$) which includes the detector positioning.

In order to set the cuts used for the special event selection, we have analysed the data taken from December 2007 to May 2008 
(around 109 active days) with a 10 line detector. During this period, around 350 up-going neutrino candidates were 
recorded. The figure 1 shows the elevation distribution of the well reconstructed muon events which pass the quality cuts 
during this period. This plot shows as the same time the down-going atmospheric muons and the up-going neutrino candidates. 
It also illustrates the very low contamination of the bad reconstructed atmospheric muons in the up-going sample.

\begin{figure}[ht]
\begin{center}
\includegraphics [width=0.6\textwidth]{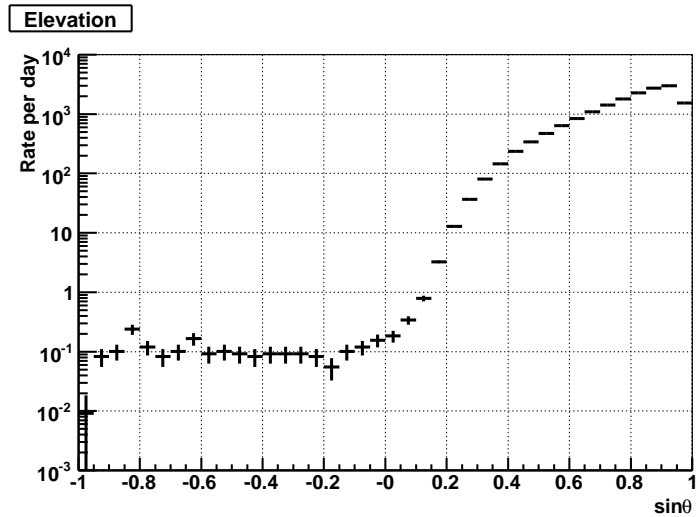}
\caption{Elevation distribution of the well reconstructed muon tracks recorded with the 10 line detector (Dec 2007 to May 2008). 
The region where the elevation is negative represents the up-going part of the distribution.}
\label{fig1}
\end{center}
\end{figure}

In order to obtain an angular resolution lower than the field of view of the telescope (around 1.9\,$^\circ$), we select reconstructed 
events which trigger several hits on at least 3 lines. The dependence of this resolution with the number of lines used in 
the fit is shown in the figure 2. For the high energy events, this resolution can be as good as 0.5 degree. 

\begin{figure}[ht]
\begin{center}
\includegraphics [width=0.6\textwidth]{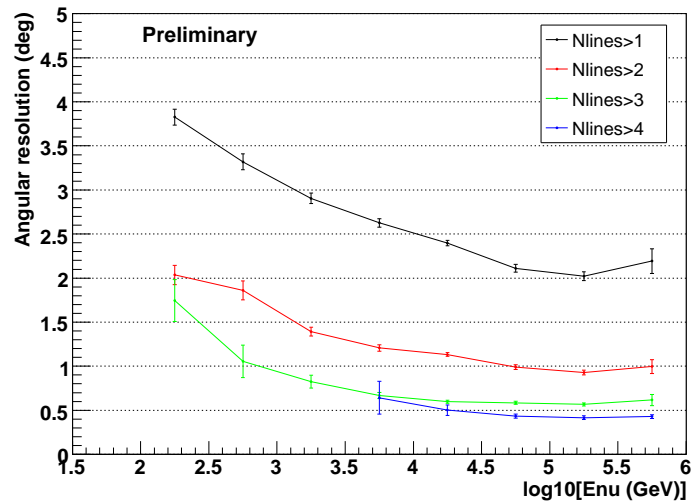}
\caption{Angular resolution evolution with energy for the event with at least 2, 3, 4 and 5 lines used in the fit.}
\label{fig2}
\end{center}
\end{figure}

An estimation of the energy in the on-line reconstruction is indirectly determined using the number of hits of the event 
and the total amplitude of these hits. In order to select events with an energy above a few tens of TeV, a minimum of about 20 hits 
and about 200 photoelectrons per track are required. These two different trigger logics applied on the 10 line data period select 
six events in about 109 active days.

%*************************************************************************************************************************
%*************************************************************************************************************************
%*************************************************************************************************************************

\section{Optical follow-up network}
\label{section3}
ANTARES is organizing a follow-up program in collaboration with TAROT (\textit{T\'elescope \`a Action Rapide pour les Objets 
Transitoires}, Rapid Action Telescope for Transient Objects; [{\cite{BTAROT}}]). This network is composed of two 25\,cm 
optical telescopes located at Calern (South of France) and La Silla (Chile). The main advantages of the TAROT instruments 
are the large field of view of 1.86\,$^\circ$ x 1.86\,$^\circ$ and their very fast positioning time (less than 10\,s). These
 telescopes are perfectly tailored for such a program. Since 2004, they observe automatically the alerts provided 
by different GRB satellites [{\cite{BTAROTgrb}}]. 

As it was said in the section 2, the rolling search method is sensitive to all transient sources which produced 
high energy neutrinos. Different observation strategies will be defined according to the different source 
types. For example a GRB afterglow requires a very fast observation strategy in contrary to a core collapse supernovae 
for which the signal will appear several days after the neutrino signal.

Such a program would not require a large observation time. Depending on the neutrino trigger settings, an alert sent to TAROT by this 
rolling search program would be issued at a rate of about one or two times per month.

%*************************************************************************************************************************
%*************************************************************************************************************************
%*************************************************************************************************************************

\section{Summary}
\label{section4}

The detection of neutrinos from transient sources is favorized by the fact that external triggers are provided by 
spacecraft currently in operation. The follow-up of interesting events would improve significantly the perspective 
for neutrino detection from transient sources. These special events are selected with two complementary triggers: 
search of a neutrino doublet (the most sensitive) or a single high energy event. The most important point of the 
rolling search method is that it is sensitive to any transient source. The confirmation by the optical telescope of a 
neutrino alert will give not only the nature of the source but also allow to increase the precision of the source 
direction determination in order to target other observatories (for example very large telescopes for the redshift 
measurement).

The key of the success of the search method is to analyze all the events on a very fast way while keeping a good 
angular precision. This good angular resolution will be an advantage not only to reduce the background and also to 
increase the probability to find an optical counterpart. The ANTARES Collaboration has an agreement with TAROT to 
develop this Target-of-Opportunity program. The implementation of this new technique has already started and we 
expect to send the first alert by the end of 2008.

%%-----------------------------
%%      your bibliography
%%-----------------------------
%In preparing the reference list please adhere to the following format.
% Attention should be paid to the order of the items in each reference
% and to the punctuation used. Please see Sect. 4 in the User's Guide
% that comes with the new macro package.

\end{document}